\documentclass[aps,prl,
 amsmath,amssymb,
 twocolumn, groupedaddress,
 showpacs]{revtex4-1}
\usepackage{bm}
\usepackage{graphicx}
\usepackage{gensymb}
\usepackage{amsmath}
\usepackage{amssymb}
\usepackage{color}
\usepackage{ulem}

\begin{document}

\title{Three-dimensional band structure of LaSb and CeSb:\\
Absence of band inversion}

\author{H. Oinuma,$^1$ S. Souma,$^{2,3}$ D. Takane,$^1$ T. Nakamura,$^1$ K. Nakayama,$^1$ 
 T. Mitsuhashi,$^{1,4}$ \\
 K. Horiba,$^4$ H. Kumigashira,$^{1,4}$ M. Yoshida,$^5$ A. Ochiai,$^1$ T. Takahashi,$^{1,2,3}$ and T. Sato$^{1,2}$}
 
\affiliation{$^1$Department of Physics, Tohoku University, Sendai 980-8578, Japan\\
$^2$Center for Spintronics Research Network, Tohoku University, Sendai 980-8577, Japan\\
$^3$WPI Research Center, Advanced Institute for Materials Research, Tohoku University, Sendai 980-8577, Japan\\
$^4$Institute of Materials Structure Science, High Energy Accelerator Research Organization (KEK), Tsukuba, Ibaraki 305-0801, Japan\\
$^5$Max-Planck-Institute for Solid State Research, Heisenbergstrsse 1, 70569 Stuttgart, Germany
}

\date{\today}

\begin{abstract}
    We have performed angle-resolved photoemission spectroscopy (ARPES) of LaSb and CeSb, a candidate of topological insulator. Using soft-x-ray photons, we have accurately determined the three-dimensional bulk band structure and revealed that the band inversion at the Brillouin-zone corner - a prerequisite for realizing topological-insulator phase - is absent in both LaSb and CeSb. Moreover, unlike the ARPES data obtained with soft-x-ray photons, those with vacuum ultraviolet (VUV) photons were found to suffer significant $k_z$ broadening. These results suggest that LaSb and CeSb are topologically trivial semimetals, and unusual Dirac-cone-like states observed with VUV photons are not of the topological origin.

\end{abstract}

\pacs{71.20.-b, 73.20.At, 79.60.-i}

\maketitle

Topological insulators (TIs) manifest a novel quantum state of matter where gapless edge or surface states (SSs) appear within inverted bulk-band gap induced by strong spin-orbit coupling \cite{ZhangReview, HasanReview, AndoReview}. The SSs of three-dimensional (3D) TIs are characterized by the linearly dispersive Dirac-cone energy band which can be viewed as a sea of Dirac fermions whose massless nature is guaranteed by the time-reversal symmetry. The discovery of TIs ignited the search for Dirac fermions associated with other symmetries, as represented by the discovery of topological crystalline insulators \cite{HsiehNC2012, TanakaNP2012, XuNC2012, DziawaNM2012}, Dirac/Weyl semimetals \cite{NeupaneNC2014, BoriPRL2014, YLChenSci2014, XuSci2015, BQLuPRX2015, YLChenNP2015, SoumaPRB2016}, and line-node semimetals \cite{SchoopNC2016, TakanePRB2016}.

Recently, it was predicted from the first-principles band-structure calculations by Zeng {\it et al.} \cite{ZengArxiv2015} that lanthanum (La) monopnictide with rock-salt structure La{\it X} ({\it X} = N, P, As, Sb, and Bi) becomes either topological Dirac semimetal (for {\it X} = N) or 3D TI (for {\it X} = P, As, Sb, and Bi) with Dirac fermions at the surface, due to the band inversion at the X point of bulk fcc Brillouin zone (BZ). While rare-earth monopnictide {\it RX} ({\it R}: rare earth) was intensively studied in 1980-90's in relation to heavy-fermion physics \cite{Boehm1979, Rossat1985, Kasuya1985, dHvA1, dHvA2, CeSbKumi, LaSbKumi}, the proposal by Zeng {\it et al.} \cite{ZengArxiv2015} renewed the interest for {\it RX} in topological aspects and triggered intensive transport, spectroscopic, and theoretical studies \cite{GuoPRB2016, GuoArxiv2016, GuoarXiv2017, Stepanov2015, TaftiNP2016, SunNJP2016, KumarPRB2016, TaftiPNAS2016, KumarArXiv2017, TaftiPRB2017, SinghaarXiv2017, YearXiv2017, ZengPRL2016, WuPRB2016, HasanArxiv2016, NiuPRB2016, NayakNC2017, LouPRB2017, WuarXiv2017}, resulting in the discovery of extremely large magnetoresistance and unusual resistivity plateau in LaSb and LaBi \cite{TaftiNP2016, TaftiPNAS2016, SunNJP2016, KumarPRB2016, TaftiPRB2017, SinghaarXiv2017, YearXiv2017}, as well as the observation of Dirac-cone-like feature in some {\it RX} compounds \cite{NiuPRB2016, WuarXiv2017, LouPRB2017, NayakNC2017, WuPRB2016, HasanArxiv2016, WuarXiv2017}.

Specifically, angle-resolved photoemission spectroscopy (ARPES) experiments of LaBi \cite{NiuPRB2016, LouPRB2017, NayakNC2017, WuPRB2016} have identified unusual Dirac-cone states at $\bar{\rm M}$ [equivalent to the projected bulk X point; see Fig. 1(a)] and $\bar{\Gamma}$ of the surface BZ, pointing to the topologically non-trivial nature due to the band inversion at X (however, note that some differences in the number and massless/massive nature of the Dirac cones are observed among different groups \cite{NiuPRB2016, LouPRB2017, NayakNC2017, WuPRB2016}). On the other hand, the electronic structure of LaSb is highly controversial. The ARPES study with vacuum ultraviolet (VUV) photons by Zeng {\it et al.} suggested the topologically {\it trivial} nature by reporting the absence of band inversion \cite{ZengPRL2016}, while the VUV-ARPES study by Niu {\it et al.} proposed the topologically {\it non-trivial} nature by reporting a Dirac-cone-like feature \cite{NiuPRB2016}. Another VUV-ARPES study of CeSb by Alidoust {\it et al.} suggested the {\it non-trivial} nature by observing two sets of anisotropic Dirac-cone-like states \cite{HasanArxiv2016}.

\begin{figure*}
\begin{center}
\includegraphics[width=6.7in]{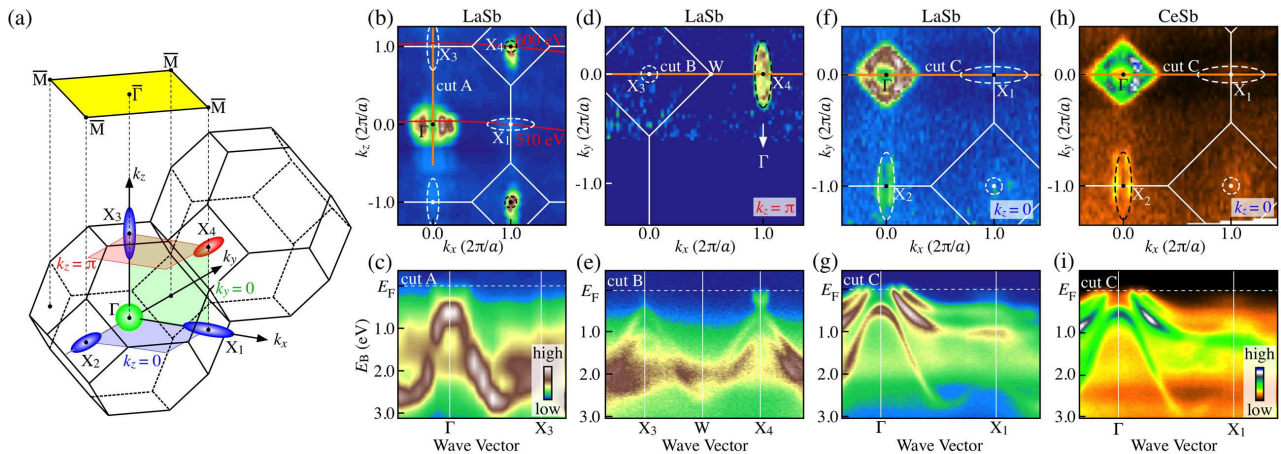}
\caption{(a) Bulk and surface BZs for LaSb and CeSb with the schematic bulk FSs at the $\Gamma$ point (green) and the X point (red and blue). For convenience, different X points are labeled as X$_1$-X$_4$. (b) ARPES-intensity mapping at $E_{\rm F}$ of LaSb as a function of $k_x$ and $k_z$, measured at $T$ = 40 K by varying $h\nu$ from 400 to 630 eV. (c) ARPES-intensity plot of LaSb as a function of wave vector and $E_{\rm B}$, measured along cut A ($\Gamma$X$_3$ cut) in (b). (d) ARPES-intensity mapping at $E_{\rm F}$ of LaSb as a function of in-plane wave vector ($k_x$ and $k_y$) at $k_z$ = $\pi$ ($h\nu$ = 600 eV). (e) ARPES intensity of LaSb measured along cut B (X$_3$WX$_4$ cut) in (d). (f) Same as (d) but at $k_z$ = 0 ($h\nu$  = 510 eV). (g) Same as (e) but along cut C ($\Gamma$X$_1$ cut) in (f). (h) and (i) Same as (f) and (g) but for CeSb.}
\vspace{-0.3in}
 \end{center}
 \end{figure*}
 
Controversy regarding the topological vs non-topological characteristics is also found in the first-principles calculations of LaSb \cite{GuoPRB2016}; The generalized gradient approximation (GGA) calculation with Perdew-Burke-Ernzerhof (PBE) formulas \cite{PBE1996} shows the band inversion, while the meta-GGA calculation with modified Becke-Johnson (MBJ) exchange potential \cite{MBJ2006, MBJ2009} supports no inversion. These controversies in both experiment and theory seriously hinder further understanding on the relationship between the electronic structure and the observed unusual physical properties in LaSb and CeSb. It is thus urgently required to experimentally establish the presence/absence of band inversion at X - a prerequisite for realizing the TIs.

In this paper, we report soft-x-ray (SX) ARPES study of LaSb and CeSb. By utilizing the energy tunable bulk-sensitive SX photons, we are able to overcome the experimental problems that previous VUV-ARPES experiments encountered, {\it i.e.} the surface effect and the large $k_z$ (momentum perpendicular to the surface) broadening  \cite{LaSbKumi, ZengPRL2016, NiuPRB2016, LouPRB2017}. We unambiguously demonstrate the absence of the band inversion in both LaSb and CeSb, by carefully tracing the bulk-band dispersion in whole 3D BZ. Our results suggest the non-topological nature of LaSb and CeSb, and lay foundation for understanding their unusual transport properties.

LaSb and CeSb single crystals were grown by the Bridgman method with a tungsten heater furnace. High-purity starting materials of La/Ce (4N) and Sb (6N) with the ratio of 1:1.005 were sealed in a tungsten crucible using an electron beam welder. The crucible was heated above their melting points and then slowly pulled down from the heater. The obtained crystals were characterized by the x-ray diffraction measurements. We could obtain  high-quality single crystals such as LaSb crystal with the residual resistivity ratio of 490 \cite{YoshidaJPSJ2002}. SX-ARPES measurements were performed with a Scienta-Omicron SES2002 electron analyzer with energy-tunable synchrotron light at BL2 (Musashi) in Photon Factory (PF), KEK. We used linearly polarized light (horizontal polarization) of 400-630 eV. VUV-ARPES measurements were performed at BL2 and BL28 with 35-100 eV photons. The energy resolutions for SX- and VUV-ARPES measurements were set at 150 meV and 10-30 meV, respectively. Samples were cleaved {\it in situ} in an ultrahigh vacuum of $\sim$ 1$\times$10$^{-10}$ Torr along the (100) crystal plane. Sample temperature was kept at $T$ = 40 K during measurements. The Fermi level ($E_{\rm F}$) of samples was referenced to that of a gold film evaporated onto the sample holder.

First, we present the overall band structure and Fermi-surface (FS) topology of LaSb obtained with SX photons. Figure 1(b) shows the ARPES-intensity mapping at $E_{\rm F}$ of LaSb as a function of $k_x$ and $k_z$ [$\Gamma$X$_1$X$_4$X$_3$ plane (area enclosed by green shading) in Fig. 1(a)] measured by varying $h\nu$ from 430 to 600 eV. One can clearly identify a bright intensity spot centered at the $\Gamma$ point, together with small weak spots at the X points. The intensity of these spots exhibits a strong $k_z$ dependence, as also seen from the band-structure plot along the $\Gamma$X$_3$ cut ($k_z$ axis) in Fig. 1(c); this demonstrates the bulk origin of these FSs. When we fix the photon energy at $h\nu$ = 600 eV and map out the ARPES intensity as a function of in-plane wave vector ($k_x$, $k_y$) at $k_z$ = $\pi$ plane [red shading in Fig. 1(a)] as displayed in Fig. 1(d), the small pocket at X (see X$_4$) is elongated along the $\Gamma$X direction. Such elongated nature is also visible in the ARPES-intensity mapping at $k_z$ = 0 plane [blue shading in Fig. 1(a)] in Fig. 1(f), although the intensity is weaker due to the photoelectron matrix-element effect. We found that the elongated FS is an electron pocket, since a shallow electronlike band crosses $E_{\rm F}$  around the X$_4$ point of the X$_3$WX$_4$ cut [cut B in (d)] in Fig. 1(e). In fact, when we overlaid the electron pocket estimated from the de Haas-van Alphen experiments in Figs. 1(b), 1(d), and 1(f) (dashed curves) \cite{dHvA1, dHvA2}, one can see a reasonable overlap with the ARPES intensity at each X points. 

As shown in the ARPES intensity along the  $\Gamma$X$_1$ cut [cut C in Fig. 1(f)] in Fig. 1(g), there exist two holelike bands crossing $E_{\rm F}$ around $\Gamma$, which form two hole pockets \cite{LaSbKumi, ZengPRL2016, NiuPRB2016}. According to the band calculations, these hole pockets are attributed to the topmost bulk valence band (VB) with the Sb 5$p$ orbital character, while the electron pocket at X is assigned to the lowest bulk conduction band (CB) with the La 5$d$ character \cite{Hasegawa1985, ZengArxiv2015, GuoPRB2016, GuoarXiv2017}. The present SX-ARPES result thus provides an evidence for the semimetallic nature of LaSb with holes at $\Gamma$ and electrons at X. We have also carried out SX-ARPES measurements on CeSb, and found that the overall FS topology and the band dispersion are very similar to those of LaSb, as seen from a side-by-side comparison of the FS [Figs. 1(f) and 1(h)] and the band dispersion [Figs. 1(g) and 1(i)] (note that the localized Ce 4$f$ states \cite{HasanArxiv2016} are not clearly resolved with this photon energy due to the low photoionization cross-section).

\begin{figure}
\begin{center}
 \includegraphics[width=3.2in]{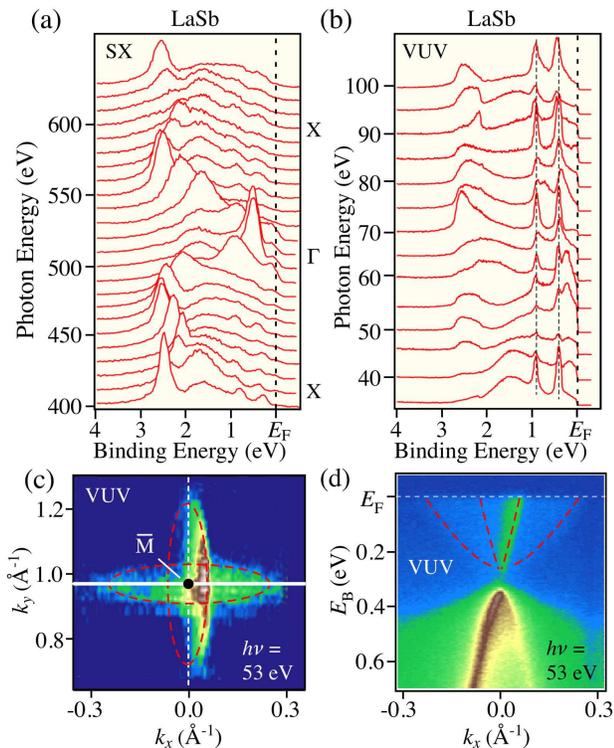}
\caption{(a) and (b) Normal-emission EDCs of LaSb measured with SX and VUV photons, respectively. (c) ARPES intensity at $E_{\rm F}$ around the $\bar{\rm M}$ point as a function of in-plane wave vector ($k_x$ and $k_y$) measured with VUV photons ($h\nu$  = 53 eV). (d) ARPES intensity as a function of $k_x$ and $E_{\rm B}$ measured with $h\nu$  = 53 eV along a cut shown by white solid line in (c). Red dashed curves in (c) and (d) are a guide for the eyes to trace FS and band dispersions.}
\vspace{-0.3in}
\end{center}
\end{figure}

We compare in Figs. 2(a) and 2(b) the normal-emission energy distribution curves (EDCs) measured with SX ($h\nu$ = 400-630 eV) and VUV photons ($h\nu$ = 35-100 eV), respectively. One can recognize several dispersive peaks in the SX data. For instance, a prominent peak at $E_{\rm B}$ $\sim$ 0.5 eV around $\Gamma$ ($h\nu$ $\sim$ 510 eV) rapidly disperses toward higher $E_{\rm B}$ upon moving away from $\Gamma$, staying at  $\sim$ 2.5 eV at midway between $\Gamma$ and X ($h\nu$ $\sim$ 450 and 550 eV), and then gradually moves back again toward 2 eV on approaching X ($h\nu$ = 600 and 410 eV). On the other hand, in the VUV data shown in Fig. 2(b), the main peaks at 0.5 and 1.0 eV show no dispersion (highlighted by dashed lines). Such features may appear owing to the  sizable $k_z$ broadening due to the short escape depth of photoelectrons excited by VUV photons \cite{LaSbKumi, NiuPRB2016, ZengPRL2016}. Taking into account the fact that the energy position of these peaks shows a good agreement with that of relatively flat bands around the X$_1$ point in the SX data [Fig. 1(g)], the $k_z$ broadening would cover a large portion of the $\Gamma$-X interval. This conclusion is also supported by the FS mapping at $h\nu$ = 53 eV in Fig. 2(c) where two pockets originating from the X$_1$- and X$_4$-centered FSs with different $k_z$'s [see Fig. 1(a)] are simultaneously observed. As a result, when we measure the band dispersion along $k_x$ at $h\nu$  = 53 eV, we accidentally observe two different electronlike bands with small and large Fermi velocities which cross the major axis of  X$_1$-centered pocket and the minor axis of  X$_4$-centered pocket, respectively, as seen in Fig. 2(d). All these results strongly suggest that the $k_z$ broadening should be carefully taken into account  when discussing the origin of $h\nu$-independent spectral feature in terms of the SS. Thus, to discuss genuine bulk electronic states free from the surface effects, it is more desirable to use SX photons.

\begin{figure}
\includegraphics[width=3.2in]{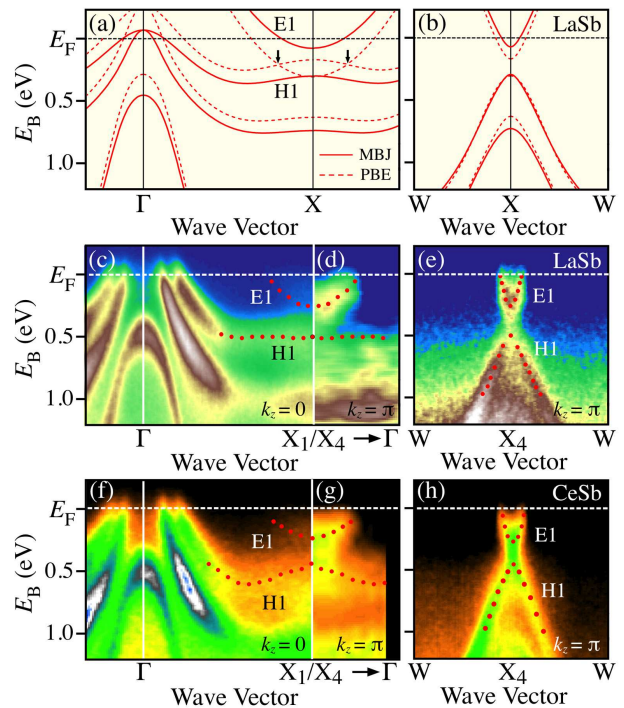}
\caption{(a) and (b) Calculated band structure for LaSb obtained with GGA calculation with PBE functional (dashed curves) and meta-GGA calculation with MBJ potential (solid curves) for $\Gamma$X and WXW high-symmetry lines, respectively \cite{GuoPRB2016}. (c) and (d) ARPES-intensity plots of LaSb measured along the $\Gamma$X$_1$ cut ($k_z$ = 0) and the $\Gamma$X$_4$ cut ($k_z$ = $\pi$), respectively (e) ARPES-intensity plot of LaSb along the WX$_4$W cut ($k_z$ = $\pi$). (f)-(h) Same as (c)-(e) but for CeSb. Red dots in (c)-(h) indicate the experimental band dispersions estimated by tracing the peak position in EDCs.}
\end{figure}

Now we turn our attention to the SX-ARPES data again to examine the possible bulk-band inversion at X. As shown in Fig. 3(a), the GGA calculation with PBE formulas for LaSb (called PBE calculation; dashed curves) \cite{GuoPRB2016} predicts that the electronlike La-5$d$ CB and the Sb-5$p$ VB (called E1 and H1 bands, respectively) intersect each other at a point slightly away from X along the $\Gamma$X cut (marked by arrow) due to the band inversion (note that a tiny spin-orbit gap opens at the intersection though it is negligible in the present energy scale). Such a band crossing is absent in the meta-GGA calculation with MBJ exchange potential (called MBJ calculation; solid curves) \cite{GuoPRB2016} since the bottom of E1 band is located at $\sim$ 0.2 eV above the H1 band at the X point. From these considerations, it can be said that a key to examine the band inversion is to experimentally verify whether such a band crossing exists or not in the SX-ARPES data.

 Figures 3(c) and 3(d) show the SX-ARPES-intensity mapping along two different $\Gamma$X cuts, $\Gamma$X$_1$ ($k_z$ = 0) and $\Gamma$X$_4$ ($k_z$ = $\pi$), respectively. We selected these two $\Gamma$X cuts because the intensity ratio between H1 and E1 bands was found to strongly depend on the choice of $k$ cut due to the afore-mentioned photoelectron matrix-element effect. In our experimental geometry, the intensity of E1 band is dramatically reduced relative to that of H1 band for $\Gamma$X$_1$ cut [Fig. 3(c)], while {\it vice versa} for $\Gamma$X$_4$ cut [Fig. 3(d)]. Therefore, in order to trace accurately the band dispersions of both H1 and E1 bands, we performed SX-ARPES measurements in these two $\Gamma$X cuts. As seen in Fig. 3(c), the H1 band exhibits a nearly flat dispersion at $E_{\rm B}$ $\sim$ 0.5 eV around the X point. On the other hand, the E1 band in Fig. 3(d) does not disperse across the H1 band with its bottom at $E_{\rm B}$ $\sim$ 0.25-0.3 eV at X. This suggests that the E1 and H1 bands do not cross each other with always keeping the energy interval of greater than 0.2 eV irrespective of $k$, in good agreement with the MBJ calculation but not with the PBE one [Fig. 3(a)]. As shown in Fig. 3(e), the absence of band crossing is also confirmed in the cut perpendicular to $\Gamma$X (WXW cut) [see also band calculation in Fig. 3(b)]. Moreover, as highlighted in Figs. 3(f)-3(h), we found that CeSb shares qualitatively the same spectral feature regarding the absence of band crossing, while the energy position and the shape of E1 and H1 bands appear to be slightly different between the two compounds. All these results strongly suggest that the bulk-band inversion does not take place in LaSb or CeSb. 

The absence of band inversion casts doubt on the TI nature proposed for LaSb and CeSb. The observed extremely large magnetoresistance in LaSb is well explained in terms of the compensation of electron and holes due to the semimetallic nature \cite{TaftiPNAS2016, ZengPRL2016, GuoPRB2016}. In addition, the observed similarity in the electronic structure between LaSb and CeSb suggests that the strong correlation of Ce 4$f$ electrons does not play a role in band inversion or possible TI characteristics. It is thus inferred that the unusual Dirac-cone-like SSs observed in the previous VUV-ARPES studies of LaSb and CeSb \cite{NiuPRB2016, HasanArxiv2016, WuarXiv2017} may be interpreted in terms of the trivial SS. Our SX-ARPES result, however, does not exclude the possibility of TI nature of LaBi and CeBi, since the band calculations of LaBi commonly predict a larger overlap of H1 and E1 bands to induce the band inversion \cite{GuoPRB2016, ZengArxiv2015, Hasegawa1985}, as supported by several previous VUV-ARPES experiments \cite{WuPRB2016, NiuPRB2016, NayakNC2017, LouPRB2017}. In this regard, the next important challenge is to perform a bulk-sensitive SX-ARPES experiment on LaBi/CeBi and clarify similarities and differences in the bulk electronic structure between LaSb and CeSb.

In summary, we performed ARPES experiments on LaSb and CeSb with bulk-sensitive soft-x-ray photons, and determined the 3D band structure and the Fermi surface. In both LaSb and CeSb, we observed two hole pockets and one electron pocket at the BZ center and corner, respectively, supporting the bulk semimetallic nature. In addition, we observed that the electronlike CB and the topmost VB are well separated around the X point, signifying the absence of bulk-band inversion. These results strongly suggest that neither LaSb or CeSb are a TI, and the unusual Dirac-cone-like energy band observed with VUV-ARPES should be associated with the topologically trivial surface states.

\begin{acknowledgements}
We thank N. Inami and K. Ono for their assistance in the ARPES measurements. This work was supported by Grant-in-Aid for Scientific Research on Innovative Areas ``Topological Materials Science'' (JSPS KAKENHI No: JP15H05853), Grant-in-Aid for Scientific Research (JSPS KAKENHI No: JP17H01139, JP15H02105, JP26287071, and JP25287079), KEK-PF (Proposal No: 2015S2-003 and 2016G555), and UVSOR (Proposal No: 28-542 and 28-828).
\end{acknowledgements}


\bibliographystyle{prsty}

\begin{thebibliography}{30}

\bibitem{HasanReview} M. Z. Hasan and C. L. Kane, Rev. Mod. Phys. \textbf{82}, 3045 (2010).

\bibitem{ZhangReview} X.-L. Qi and S.-C. Zhang, Rev. Mod. Phys. \textbf{83}, 1057 (2011).

\bibitem{AndoReview} Y. Ando, J. Phys. Soc. Jpn.  \textbf{82}, 102001 (2013).

\bibitem{HsiehNC2012} T. H. Hsieh, H. Lin, J. Liu, W. Duan, A. Bansil, and L. Fu, Nat. Commun. \textbf{3}, 982 (2012).
%
\bibitem{TanakaNP2012} Y. Tanaka, Z. Ren, T. Sato, K. Nakayama, S. Souma, T. Takahashi, K. Segawa, and Y. Ando, Nat. Phys. \textbf{8}, 800 (2012).

\bibitem{XuNC2012} S.-Y. Xu, C. Liu, N. Alidoust, M. Neupane, D. Qian, I. Belopolski, J. D. Denlinger, Y. J. Wang, H. Lin, L. A. Wray, G. Landolt, B. Slomski, J. H. Dil, A. Marcinkova, E. Morosan, Q. Gibson, R. Sankar, F. C. Chou, R. J. Cava, A. Bansil, and M. Z. Hasan, Nat. Commun. \textbf{3}, 1192 (2012).

\bibitem{DziawaNM2012} P. Dziawa, B. J. Kowalski, K. Dybko, R. Buczko, A. Szczerbakow, M. Szot, E. \L usakowska, T. Balasubramanian, B. M. Wojek, M. H. Berntsen, O. Tjernberg, and T. Story, Nat. Mater. \textbf{11}, 1023 (2012).

\bibitem{NeupaneNC2014} M. Neupane, S.-Y. Xu, R. Sankar, N. Alidoust, G. Bian, C. Liu, I. Belopolski, T.-R. Chang, H.-T. Jeng, H. Lin, A. Bansil, F. Chou, and M. Z. Hasan, Nat. Commun. \textbf{5}, 3786 (2014). 
%
\bibitem{BoriPRL2014} S. Borisenko, Q. Gibson, D. Evtushinsky, V. Zabolotnyy, B. B\"{u}chner, and R. J. Cava, Phys. Rev. Lett. \textbf{113}, 027603 (2014).

\bibitem{YLChenSci2014} Z. K. Liu, B. Zhou, Y. Zhang, Z. J. Wang, H. M. Weng, D. Prabhakaran, S.-K. Mo, Z. X. Shen, Z. Fang, X. Dai, Z. Hussain, and Y. L. Chen, Science \textbf{343}, 864 (2014).

\bibitem{XuSci2015} S.-Y. Xu, I. Belopolski, N. Alidoust, M. Neupane, G. Bian, C. Zhang, R. Sankar, G. Chang, Z. Yuan, C.-C. Lee, S.-M. Huang, H. Zheng, J. Ma, D. S. Sanchez, B. K. Wang, A. Bansil, F. Chou, P. P. Shibayev, H. Lin, S. Jia, and M. Z. Hasan, Science \textbf{349}, 613 (2015). 

\bibitem{BQLuPRX2015} B. Q. Lv, H. M. Weng, B. B. Fu, X. P. Wang, H. Miao, J. Ma, P. Richard, X. C. Huang, L. X. Zhao, G. F. Chen, Z. Fang, X. Dai, T. Qian, and H. Ding, Phys. Rev. X \textbf{5}, 031013 (2015).
%
\bibitem{YLChenNP2015} L. X. Yang, Z. K. Liu, Y. Sun, H. Peng, H. F. Yang, T. Zhang, B. Zhou, Y. Zhang, Y. F. Guo, M. Rahn, D. Prabhakaran, Z. Hussain, S.-K. Mo, C. Felser, B. Yan, and Y. L. Chen, Nat. Phys. \textbf{11}, 728 (2015). 

\bibitem{SoumaPRB2016} S. Souma, Z. Wang, H. Kotaka, T. Sato, K. Nakayama, Y. Tanaka, H. Kimizuka, T. Takahashi, K. Yamauchi, T. Oguchi, K. Segawa, and Y. Ando, Phys. Rev. B \textbf{93}, 161112(R) (2016).

\bibitem{SchoopNC2016} L. M. Schoop, M. N. Ali, C. Stra{\ss}er, A. Topp, A. Varykhalov, D. Marchenko, V. Duppel, S. S. P. Parkin, B. V. Lotsch, and C. R. Ast, Nat. Commun. \textbf{7}, 11696 (2016).

\bibitem{TakanePRB2016} D. Takane, Z. Wang, S. Souma, K. Nakayama, C. X. Trang, T. Sato, T. Takahashi, and Y. Ando, Phys. Rev. B \textbf{94}, 121108(R) (2016).

\bibitem{ZengArxiv2015} M. Zeng, C. Fang, G. Chang, Y.-A. Chen, T. Hsieh, A. Bansil, H. Lin, and L. Fu, arXiv:1504.03492 (2015).

\bibitem{Boehm1979} J. von Boehm and P. Bak, Phys. Rev. Lett. \textbf{42}, 122 (1979).
%
\bibitem{Rossat1985} J. Rossat-Mignod, J. M. Effantin, P. Burlet, T. Chattopadhyay, L. P. Regnault, H. Bartholin, C. Vettier, O. Vogt, D. Ravot, and J. C. Achart, J. Magn. Magn. Mater. \textbf{52}, 111 (1985).

\bibitem{dHvA1} H. Kitazawa, T. Suzuki, M. Sera, I. Oguro, A. Yanase, A. Hasegawa, and T. Kasuya, J. Magn. Magn. Mater. \textbf{31-34}, 421 (1983).

\bibitem{Kasuya1985} H. Takahashi and T. Kasuya, J. Phys. C \textbf{18}, 2697 (1985).

\bibitem{dHvA2} R. Settai, T. Goto, S. Sakatsume, Y. S. Kwon, T. Suzuki, and T. Kasuya, Physica B \textbf{186-188}, 176 (1993).
%
\bibitem{CeSbKumi} H. Kumigashira, H.-D. Kim, A. Ashihara, A. Chainani, T. Yokoya, T. Takahashi, A. Uesawa, and T. Suzuki, Phys. Rev. B \textbf{56}, 13654 (1997).

\bibitem{LaSbKumi} H. Kumigashira, H.-D. Kim, T. Ito, A. Ashihara, T. Takahashi, T. Suzuki, M. Nishimura, O. Sakai, Y. Kaneta, and H. Harima, Phys. Rev. B \textbf{58}, 7675 (1998).

\bibitem{GuoPRB2016} P.-J. Guo, H.-C. Yang, B.-J. Zhang, K. Liu, and Z.-Y. Lu, Phys. Rev. B \textbf{93}, 235142 (2016).

\bibitem{GuoArxiv2016} C. Y. Guo, C. Cao, M. Smidman, F. Wu, Y. J. Zhang, F. Steglich, F. C. Zhang, and H. Q. Yuan, arXiv:1611.02927v1 (2016).

\bibitem{GuoarXiv2017} P.-J. Guo, H.-C. Yang, K. Liu, and Z.-Y. Lu, arXiv:1706.04029v1 (2017).
%
\bibitem{Stepanov2015} N. N. Stepanov, N. V. Morozova, A. E. Kar'kin, A. V. Golubkov, and V. V. Kaminskii, Phys. Solid State \textbf{57}, 2369 (2015). 

\bibitem{TaftiNP2016} F. F. Tafti, Q. D. Gibson, S. K. Kushwaha, N. Haldolaarachchige, and R. J. Cava, Nat. Phys. \textbf{12}, 272 (2016).

\bibitem{SunNJP2016} S. S. Sun, Q. Wang, P. J. Guo, K. Liu, and H. C. Lei, New J. Phys. \textbf{18}, 082002 (2016).

\bibitem{KumarPRB2016} N. Kumar, C. Shekhar, S.-C. Wu, I. Leermakers, O. Young, U. Zeitler, B. H. Yan, and C. Felser, Phys. Rev. B \textbf{93}, 241106(R) (2016).

\bibitem{TaftiPNAS2016} F. F. Tafti, Q. D. Gibson, S. K. Kushwaha, J. W. Krizan, N. Haldolaarachchige, and R. J. Cava, Proc. Natl. Acad. Sci. \textbf{113}, E3475-E3481 (2016).
%
\bibitem{TaftiPRB2017} F. F. Tafti, M. S. Torikachvili, R. L. Stillwell, B. Baer, E. Stavrou, S. T. Weir, Y. K. Vohra, H.-Y. Yang, E. F. McDonnell, S. K. Kushwaha, Q. D. Gibson, R. J. Cava, and J. R. Jeffries, Phys. Rev. B \textbf{95}, 014507 (2017).

\bibitem{SinghaarXiv2017} R. Singha, B. Satpati, and P. Mandal, arXiv:1703.06100v1 (2017).

\bibitem{YearXiv2017} L. Ye, T. Suzuki, C. R. Wicker, and J. G. Checkelsky, arXiv:1704.04226v1 (2017).

\bibitem{KumarArXiv2017} N. Kumar, C. Shekhar, J. Klotz, J. Wosnitza, and C. Felser, arXiv:1703.02331 (2017).

\bibitem{ZengPRL2016} L.-K. Zeng, R. Lou, D.-S. Wu, Q. N. Xu, P.-J. Guo, L.-Y. Kong, Y.-G. Zhong, J.-Z. Ma, B.-B. Fu, P. Richard, P. Wang, G. T. Liu, L. Lu, Y.-B. Huang, C. Fang, S.-S. Sun, Q. Wang, L. Wang, Y.-G. Shi, H. M. Weng, H.-C. Lei, K. Liu, S.-C. Wang, T. Qian, J.-L. Luo, and H. Ding, Phys. Rev. Lett. \textbf{117}, 127204 (2016).

\bibitem{HasanArxiv2016} N. Alidoust, A. Alexandradinata, S.-Y. Xu, I. Belopolski, S. K. Kushwaha, M. Zeng, M. Neupane, G. Bian, C. Liu, D. S. Sanchez, P. P. Shibayev, H. Zheng, L. Fu, A. Bansil, H. Lin, R. J. Cava, and M. Z. Hasan, arXiv:1604.08571v1 (2016).

\bibitem{WuPRB2016} Y. Wu, T. Kong, L.-L. Wang, D. D. Johnson, D. Mou, L. Huang, B. Schrunk, S. L. Bud'ko, P. C. Canfield, and A. Kaminski, Phys. Rev. B \textbf{94}, 081108(R) (2016).

\bibitem{NiuPRB2016} X. H. Niu, D. F. Xu, Y. H. Bai, Q. Song, X. P. Shen, B. P. Xie, Z. Sun, Y. B. Huang, D. C. Peets, and D. L. Feng, Phys. Rev. B \textbf{94}, 165163 (2016).

\bibitem{NayakNC2017} J. Nayak, S.-C. Wu, N. Kumar, C. Shekhar, S. Singh, J. Fink, E. E. D. Rienks, G. H. Fecher, S. S. P. Parkin, B. Yan, and C. Felser, Nat. Commun. \textbf{8}, 13942 (2017).

\bibitem{LouPRB2017} R. Lou, B.-B. Fu, Q. N. Xu, P.-J. Guo, L.-Y. Kong, L.-K. Zeng, J.-Z. Ma, P. Richard, C. Fang, Y.-B. Huang, S.-S. Sun, Q. Wang, L. Wang, Y.-G. Shi, H. C. Lei, K. Liu, H. M. Weng, T. Qian, H. Ding, and S.-C. Wang, Phys. Rev. B \textbf{95}, 115140 (2017).
%
\bibitem{WuarXiv2017} Y. Wu, Y. Lee, T. Kong, D. Mou, R. Jiang, L. Huang, S. L. Bud'ko, P. C. Canfield, and A. Kaminski, arXiv:1704.06237v1 (2017).

\bibitem{PBE1996} J. P. Perdew, K. Burke, and M. Ernzerhof, Phys. Rev. Lett. \textbf{77}, 3865 (1996).

\bibitem{MBJ2006} A. D. Becke and E. R. Johnson, J. Chem. Phys. \textbf{124}, 221101 (2006).

\bibitem{MBJ2009} F. Tran and P. Blaha, Phys. Rev. Lett. \textbf{102}, 226401 (2009).

\bibitem{YoshidaJPSJ2002} M. Yoshida, K. Koyama, T. Tomimatsu, M. Shirakawa, A. Ochiai, and M. Motokawa, J. Phys. Soc. Jpn. \textbf{71}, 1752 (2002).

\bibitem{Hasegawa1985} A. Hasegawa, J. Phys. Soc. Jpn. \textbf{54}, 677 (1985).


 
\end{thebibliography}

\end{document}